\documentclass[12pt,notoc,nohyper]{QCEMDA}
\usepackage{amssymb}
\usepackage{epsfig,multicol,bbm}

\makeatletter
\def\blfootnote{\xdef\@thefnmark{Hi my dear}\@footnotetext}
\makeatother

\newcommand\fverb{\setbox\pippobox=\hbox\bgroup\verb}
\newcommand\fverbdo{\egroup\medskip\noindent            \fbox{\unhbox\pippobox}\ }
\newcommand\fverbit{\egroup\item[\fbox{\unhbox\pippobox}]}
\newbox\pippobox
\preprint{}

\title{Quantum tunneling from three-dimensional black holes}
\author{Asiya Ejaz$^1$, H. Gohar$^1$, Hai Lin$^{2,3}$, K. Saifullah$^{1,2,4}$%
, Shing-Tung Yau$^{2,3}$ \\
$^1$Department of Mathematics, Quaid-i-Azam University, Islamabad, Pakistan \\
$^2$Department of Physics, and Center for the Fundamental Laws of
Nature,\\
~Harvard University, MA 02138, USA \\
$^3$Department of Mathematics, Harvard University, MA 02138, USA \\
$^4$Department of Physics, University of Massachusetts, Amherst, MA 01003,
USA \\
\\
Electronic address: hailin@fas.harvard.edu,
ksaifullah@fas.harvard.edu, \\
~~~~~~~~~~~~~~~~~~~~~~~~yau@math.harvard.edu}

\abstract{We study Hawking radiation from three-dimensional black
holes. For this purpose the emission of charged scalar and charged
fermionic particles is investigated from charged BTZ black holes,
with and without rotation. We use the quantum tunneling approach
incorporating WKB approximation and spacetime symmetries. Another
class of black hole which is asymptotic to a Sol three-manifold
has also been investigated. This procedure gives us the tunneling
probability of outgoing particles, and we compute the temperature
of the radiation for these black holes. We also consider the
quantum tunneling of particles from black hole asymptotic to Sol
geometry. }

\begin{document}

\section{Introduction}

Quantum mechanical effects when studied in the background of
classical general relativity give rise to many interesting
phenomenon. These phenomena play an important role in
understanding the theories of quantum gravity. One such process of
significance is the evaporation of black holes as a result of
Hawking radiations \cite{Hawking:1974sw, Gibbons:1977mu}. These
radiations have also been viewed as quantum tunneling of particles
from the horizons of black holes \cite{Kraus:1994fh,
Parikh:1999mf, Iso:2006ut}. Many well known black holes have been
researched for these radiations [6-12]. In one of the procedures
the wave equation governing the emission of particles is solved in
the background of the black hole spacetimes by using complex path
integration techniques and WKB approximation. This gives the
tunneling probability of particles crossing the event horizon,
which in turn gives the temperature and surface gravity of the
black hole.

The study of (2+1)-dimensional black holes provides a valuable insight in
understanding low-dimensional gravity theories and their quantum
counterparts. The BTZ black hole \cite{Banados:1992wn} is such an example in
(2+1)-dimensional gravity. This black hole is remarkably similar to
(3+1)-dimensional black hole. Like the Kerr black hole it contains an inner
and an outer horizon. It can be fully characterized by mass, angular
momentum and charge. It also possesses thermodynamical properties analogous
to the (3+1)-dimensional black hole. Its entropy is given by a law analogous
to the Bekenstein bound in (3+1)-dimensions, where we replace the surface
area by the circumference of the BTZ black holes . This black hole does
arise from collapsing matter and can represent a gravitational collapse. The
BTZ solution is also discussed in the realm of (2+1)-dimensional quantum
gravity.

In this paper, first we consider the charged version of the BTZ black hole
\cite{Banados:1992wn}. The charged BTZ black hole is characterized by a
power-law curvature singularity generated by the electric charge of the
hole. The curvature singularity gives rise to $\ln r$ terms when the
gravitational field is expanded asymptotically and it has nontrivial
boundary terms. This black hole solution exists in the presence of a
negative or zero cosmological constant. In this paper we consider tunneling
of charged scalar and charged fermionic particles from these black holes,
and work out the Hawking temperature. We also study a class of topological
three-dimensional black holes constructed from Sol geometry \cite{Scott}. In
our approach we solve charged Klein-Gordon and Dirac equations and calculate
the tunneling probabilities of particles crossing their horizons and work
out the temperatures.

The organization of this paper is as follows. After an introduction of BTZ
black holes in the next section, we discuss tunneling of charged scalar
particles from charged BTZ black holes in Section \ref{sec_
charge_scalar_charge}, and from charged rotating BTZ black holes in Sections %
\ref{sec_ charge_scalar_charge_rotation}. After this, Section \ref{sec_
charge_fermionic_charge} and Section \ref{sec_
charge_fermionic_charge_rotation} are devoted to the emission of charged
fermions from charged black holes, and from charged rotating black holes,
respectively. In Section \ref{sec_ particle_sol}, we investigate the
topological black hole from the Sol geometry and work out the tunneling
probability and Hawking temperature. We conclude our paper with a discussion
and brief summary of results.

\section{Charged BTZ black holes}

\label{sec_ bh}

The BTZ black hole solutions in (2+1) spacetime dimensions are derived from
a three-dimensional theory of gravity. The BTZ black holes \cite%
{Banados:1992wn, Banados:1992gq} are solutions of the Einstein field
equations with cosmological constant in three dimensions. These solutions
represent one of the main recent advances for low energy in three
dimensional gravity theories. The total Einstein action in three dimensional
gravity in the presence of cosmological constant is given by
\begin{equation}
I=\frac{1}{16\pi G}\int d^{3}x\sqrt{-g}\left( R-2\Lambda \right) ,  \label{E}
\end{equation}%
where $G$ is the gravitational constant, $\Lambda =-1/{l^{2}}$, is the
Cosmological constant, $R$, the Ricci scalar and $g$ is determinant of the
metric tensor $g_{\mu \nu }$. We use units for both $G$ and $l$ as$~($length)%
$^{3}$ and we work in the units that $8G=1$ in this paper.~

The corresponding line element in Schwarzschild coordinates is given by
\begin{equation}
ds^{2}=-f\left( r\right) dt^{2}+f^{-1}\left( r\right) dr^{2}+r^{2}\left(
d\phi -\frac{J}{2r^{2}}dt\right) ^{2},  \label{BTZ_}
\end{equation}%
where%
\begin{equation}
f\left( r\right) =-M+\frac{r^{2}}{l^{2}}+\frac{J^{2}}{4r^{2}}.
\end{equation}%
Here $M$ is the mass, $J$ is the angular spin of the BTZ black hole and
\begin{equation}
-\infty <t<\infty ,\quad 0\leq r<\infty ,\quad 0\leq \phi <2\pi .
\end{equation}%
The horizons, $r_{+}$ $($henceforth simply the black hole horizon) and $%
r_{-} $ respectively, concerning the positive mass black hole spectrum with
spin $\left( J\neq 0\right) $ of the line element (\ref{BTZ_}) are given by
putting $f\left( r\right) =0$,
\begin{equation}
r_{\pm }^{2}=\frac{l^{2}}{2}\left( M\pm \sqrt{M^{2}-\frac{J^{2}}{l^{2}}}%
\right) ,
\end{equation}%
where `+' and `$-$' denote the outer and inner horizons. The BTZ black hole
without electric charge can be obtained as the quotient of AdS space.

One can obtain more general metrics by considering coupling of the
pure Einstein gravity with other matter fields. For example, one
can consider three-dimensional Einstein gravity with topological
matter \cite{Carlip:1991zk}. One can also discuss the
Einstein-Maxwell theory. If we include the Maxwell tensor also,
the action is given by
\begin{equation}
I=\frac{1}{16\pi G}\int d^{3}x\sqrt{-g}\left( R-2\Lambda -4\pi GF_{\mu \nu
}F^{\mu \nu }\right) ,  \label{Einstein_Maxwell}
\end{equation}%
with
\begin{equation}
F_{\mu \nu }=A_{\nu ,\mu }-A_{\mu ,\nu },
\end{equation}
where $A_{\mu }$ is the electrical potential. In addition to the black hole
solutions (\ref{BTZ_}) described above, charged black hole solutions similar
to (\ref{BTZ_}) exist \cite{Banados:1992wn, Martinez:1999qi}. These are
solutions following from the action (\ref{Einstein_Maxwell}).

The electrically charged black hole solutions of Einstein-Maxwell theory
take the form
\begin{equation}
ds^{2}=-f\left( r\right) dt^{2}+f^{-1}\left( r\right) dr^{2}+r^{2}d\phi ^{2},
\label{bh_Q}
\end{equation}
but with%
\begin{equation}
f(r)=-M+\frac{r^{2}}{l^{2}}-\frac{1}{2}Q^{2}\ln \left( \frac{r}{l}\right) ,
\label{f_r_Q_}
\end{equation}
where $Q$ is the electric charge of the black hole. For this charged black
hole, there is a limit $l\rightarrow \infty ,\Lambda \rightarrow 0,~$in
which it goes over to charged black hole asymptotic to flat space.

The electrically charged and rotating black holes take the form (\ref{BTZ_}%
), but with%
\begin{equation}
f(r)=-M+\frac{r^{2}}{l^{2}}-\frac{1}{2}Q^{2}\ln \left( \frac{r}{l}\right) +%
\frac{J^{2}}{4r^{2}},  \label{f_r_Q__}
\end{equation}%
where $Q$ is the electric charge, and $J$ is the angular momentum of the
black hole.

Using the fact that for BTZ black hole electric potential, $A_{\mu }$, is
given by $\left( A_{t},0,0\right) ,$ we have from (\ref{Einstein_Maxwell})
and (\ref{f_r_Q_})
\begin{equation}
\frac{Q}{r}=-A_{t,r}.
\end{equation}%
After integration the above expression comes out to be%
\begin{equation}
A_{t}=-Q\ln (\frac{r}{l}).
\end{equation}%
Since we are considering the case of charged BTZ black hole, we consider the
line element (\ref{bh_Q}) with $f(r)$ given by (\ref{f_r_Q_}). The Maxwell
field is given by
\begin{equation}
F_{tr}=\frac{Q}{r}.
\end{equation}

We see that unlike the uncharged case where horizons can be found
easily, the function $f(r)$ is more complicated for the charged
BTZ black hole.

%%%%%%%%%%%%%%%%%%%%%%%%%%%%%%%%%%%%%%%%%%%%%%%%%%%%%%%%%%%%%%%%%%%
%%%%%%%%%%%%%%%%%%%%%%%%%%%%%%%%%%%%%%%%%%%%%%%%%%%%%%%%%%%%%%%%%%%
%%%%%%%%%%%%%%%%%%%%%%%%%%%%%%%%%%%%%%%%%%%%%%%%%%%%%%%%%%%%%%%%%%%

\section{Quantum tunneling of scalar particles from charged BTZ black hole}

\label{sec_ charge_scalar_charge}

Here we treat the emission of scalar particles from charged BTZ black holes
as tunneling phenomenon across the event horizon. This is done by solving
charged Klein-Gordon equation for scalar field, $\Psi $, which is given by
\begin{equation}
\frac{1}{\sqrt{-g}}\left( \partial _{\mu }-\frac{iq}{\hbar }A_{\mu }\right)
\left( \sqrt{-g}g^{\mu \nu }(\partial _{\nu }-\frac{iq}{\hbar }A_{\nu })\Psi
\right) -\frac{m^{2}}{\hbar ^{2}}\Psi =0,  \label{K_G_1}
\end{equation}%
where $\nu $ and $\mu $ have values $0,1,2$ for the coordinates $t,r,\phi .$
The $m$ and $q$ are the mass and charge of the particle. We use WKB
approximation and choose an ansatz of the form
\begin{equation}
\Psi (t,r,\phi )=e^{\left( \frac{i}{\hslash }I(t,r,\phi )+I_{1}(t,r,\phi
)+O(\hslash )\right) }.  \label{wave_factor_}
\end{equation}%
Using this in (\ref{K_G_1}) in leading powers of $\hbar $, dividing by the
exponential term and multiplying by $\hbar ^{2},$ we get%
\begin{equation}
0=g^{tt}(\partial _{t}I-qA_{t})^{2}+g^{rr}(\partial _{r}I)^{2}+g^{\phi \phi
}(\partial _{\phi }I)^{2}+m^{2},
\end{equation}%
or
\begin{eqnarray}
0 &=&-\left( -M+\frac{r^{2}}{l^{2}}-\frac{1}{2}Q^{2}\ln r\right)
^{-1}(\partial _{t}I-qA_{t})^{2}+  \nonumber \\
&&\left( -M+\frac{r^{2}}{l^{2}}-\frac{1}{2}Q^{2}\ln r\right) (\partial
_{r}I)^{2}+r^{-2}(\partial _{\phi }I)^{2}+m^{2}.  \label{wave_Q_}
\end{eqnarray}

If we look at the symmetries of charged BTZ black hole then $\partial _{t}$
and $\partial _{\phi }$ are the Killing fields. So there exists a solution
for this differential equation which in terms of the classical action $I$
can be written as
\begin{equation}
I=-\omega t+W(r)+j\phi +K,  \label{phase_}
\end{equation}%
where $\omega $ and $j$ are the energy and angular momentum of the particle,
and $K$ is a constant which can be complex. Using this function in the above
expression we get
\begin{equation}
W_{\pm }(r)=\pm \int \sqrt{-\frac{g^{tt}}{g^{rr}}\left( (\omega +qA_{t})^{2}+%
\frac{g^{\phi \phi }}{g^{tt}}j^{2}+\frac{1}{g^{tt}}m^{2}\right) }dr.
\end{equation}%
Putting the values of $g^{tt}$ and $g^{rr},$ we can write $W_{\pm }(r)$ as%
\begin{equation}
W_{\pm }(r)=\pm \int \frac{\sqrt{\left( \omega +qA_{t}\right)
^{2}-f(r)\left( \frac{j^{2}}{r^{2}}+m^{2}\right) }}{f(r)}dr.
\end{equation}%
Here, we have simple pole at $r=r_{+},$ so by using the residue theory for
semi circles, we get
\begin{equation}
W_{\pm }=\pm \frac{\pi i(\omega +qA_{t})}{f^{\prime }(r_{+})},
\label{phase_r_}
\end{equation}%
since $f(r_{+})=0$. This implies that
\begin{equation}
\mathrm{Im}W_{+}=\frac{\pi (\omega +qA_{t})}{f^{\prime }(r_{+})}.
\label{phase_imaginary_}
\end{equation}

Hawking radiation from black holes can be studied as a process of quantum
tunneling of particles from the black hole horizon. Using this approach we
calculate the imaginary part of the classical action for this classically
forbidden process of emission across the horizon. In this semi-classical
method the probabilities of crossing the horizon from inside to outside, and
from outside to inside, are given by \cite{Shankaranarayanan:2000gb,
Srinivasan:1998ty}
\begin{eqnarray}
P_{emission} &\varpropto &\exp \left( \frac{-2}{\hbar }\mathrm{Im}I\right)
=\exp \left( \frac{-2}{\hbar }(\mathrm{Im}W_{+}+\mathrm{Im}K)\right) ,
\label{p_emit_} \\
P_{absorption} &\varpropto &\exp \left( \frac{-2}{\hbar }\mathrm{Im}I\right)
=\exp \left( \frac{-2}{\hbar }(\mathrm{Im}W_{-}+\mathrm{Im}K)\right) .
\label{p_absorb_}
\end{eqnarray}%
We know that any outside particle will certainly fall into the black hole.
Thus we must take Im$K=-$Im$W_{-}$. From (\ref{phase_r_}), we have $%
W_{+}=-W_{-}$, and this means that the probability of a particle tunneling
from inside to outside the horizon is
\begin{equation}
\Gamma =\exp \left( \frac{-4}{\hbar }\mathrm{Im}W_{+}\right) .
\label{rate_phase_}
\end{equation}%
Putting the value of Im$W_{+}$ from equation (\ref{phase_imaginary_}) into (%
\ref{rate_phase_}), we get
\begin{equation}
\Gamma =\exp \left( -\frac{4\pi (\omega -qQ\ln \left( \frac{r_{+}}{l}\right)
)}{\hbar f^{\prime }(r_{+})}\right) .  \label{rate_Q}
\end{equation}%
\newline
This is the tunneling probability of scalar particles from inside to outside
the event horizon of the charged BTZ black hole.

If we compare this equation (\ref{rate_Q}) with $\Gamma =\exp
\left( -\beta \omega \right) $, which is Boltzmann factor for
particle of energy $\omega$, and $\beta $ is the inverse temperature of the horizon \cite%
{Shankaranarayanan:2000gb, Srinivasan:1998ty}, we can derive the Hawking
temperature for black holes. Comparing equation (\ref{rate_Q}) with the
Boltzmann factor of energy, we can find the Hawking temperature (taking $%
\hbar =1$) as
\begin{equation}
T_{H}=\frac{f^{\prime }(r_{+})}{4\pi },  \label{T_tunneling_01}
\end{equation}%
where $f^{\prime }(r_{+})$ is the derivative of $f$ with respect to $r$ at $%
r=r_{+}.$ So from equation (\ref{T_tunneling_01}), the temperature becomes%
\begin{equation}
T_{H}=\frac{1}{4\pi }\left( \frac{2r_{+}}{l^{2}}-\frac{Q^{2}}{2r_{+}}\right)
.  \label{T_tunneling_Q}
\end{equation}%
This situation is similar to the Reissner-Nordstr\"{o}m black hole in
(3+1)-dimensions. It has the interesting Boltzmann factor (\ref{rate_Q}),
with chemical potential conjugate to the charge.

Now we look at thermodynamic relations in this situation. From
$f(r_{+})=0$, the mass of the black hole is given by
\begin{equation}
M_{bh}(Q)=\frac{r_{+}^{2}}{l^{2}}-\frac{1}{2}Q^{2}\ln \left( \frac{r_{+}}{l}%
\right) .
\end{equation}%
The entropy of the black hole is%
\begin{equation}
S=4\pi r^{+},
\end{equation}%
which is the circumference, in the units $8G=1$. For the electric
potential $ V$
\begin{equation}
\frac{\partial M}{\partial Q}|_{S}=V=-Q\ln \left( \frac{r_{+}}{l}\right)
=A_{t}.
\end{equation}%
So that we have
\begin{equation}
\frac{\partial M}{\partial S}|_{Q}=T=\frac{1}{4\pi }\left( \frac{2r_{+}}{%
l^{2}}-\frac{Q^{2}}{2r_{+}}\right) ,
\end{equation}%
which is from the thermodynamic relation, and is the same as (\ref%
{T_tunneling_Q}) by the above independent method of quantum tunneling.

%%%%%%%%%%%%%%%%%%%%%%%%%%%%%%%%%%%%%%%%%%%%%%%%%%%%%%%%%%%%%%%%%%%
%%%%%%%%%%%%%%%%%%%%%%%%%%%%%%%%%%%%%%%%%%%%%%%%%%%%%%%%%%%%%%%%%%%
%%%%%%%%%%%%%%%%%%%%%%%%%%%%%%%%%%%%%%%%%%%%%%%%%%%%%%%%%%%%%%%%%%%

\section{Quantum tunneling of scalar particles from rotating charged BTZ
black hole}

\label{sec_ charge_scalar_charge_rotation}

In this section, we consider emission of scalar particles from rotating
charged BTZ black hole. The line element is given by%
\begin{equation}
ds^{2}=-f\left( r\right) dt^{2}+f^{-1}\left( r\right) dr^{2}+r^{2}\left(
d\phi +N^{\phi }(r)dt\right) ^{2},
\end{equation}%
where%
\begin{eqnarray}
f(r) &=&-M+\frac{r^{2}}{l^{2}}+\frac{J^{2}}{4r^{2}}-\frac{1}{2}Q^{2}\ln
\left( \frac{r}{l}\right) , \\
N^{\phi }(r) &=&\frac{-J}{2r^{2}}.
\end{eqnarray}
To deal with the quantum tunneling of scalar particles from this black hole,
we will use the charged Klein-Gordon equation given by (\ref{K_G_1}). Again
assuming the function of the form in (\ref{wave_factor_}) for the solution
and following the earlier procedure, we get the differential equation%
\begin{equation}
g^{tt}(\partial _{t}I-qA_{t})^{2}+g^{rr}(\partial _{r}I)^{2}+g^{t\phi
}(\partial _{t}I\partial _{\phi }I-qA_{t}\partial _{\phi }I)+g^{\phi \phi
}(\partial _{\phi }I)^{2}+m^{2}=0.
\end{equation}%
As before we assume the function $I$ of the form in (3.6) and obtain
\begin{equation}
W^{\prime }(r)=\pm \sqrt{-\frac{g^{tt}}{g^{rr}}\left( (\omega +qA_{t})^{2}+%
\frac{g^{\phi \phi }}{g^{tt}}j^{2}-\frac{g^{t\phi }}{g^{tt}}j(\omega
+qA_{t})+\frac{1}{g^{tt}}m^{2}\right) }.
\end{equation}%
Substituting the value of the metric tensor we get the integral
\begin{equation}
W_{\pm }(r)=\pm \int \frac{\sqrt{(\omega +qA_{t})^{2}-(\frac{f(r)}{r^{2}}%
-\left( N^{\phi }\right) ^{2})j^{2}+2N^{\phi }(r)(\omega +qA_{t})j-f(r)m^{2}}%
}{f(r)}dr.
\end{equation}
Here, we have simple pole at $r=r_{+}$, and therefore, from the residue
theory for semi circles, we get on integration
\begin{equation}
W_{\pm }=\pm \pi i\frac{\sqrt{(\omega +qA_{t})^{2}+\left( N^{\phi
}(r_{+})\right) ^{2}j^{2}+2N^{\phi }(r_{+})(\omega +qA_{t})j}}{f^{\prime
}(r_{+})}.
\end{equation}
The above equation implies that
\begin{equation}
\mathrm{Im}W_{+}=\pi \frac{\left( \omega +qA_{t}(r_{+})+jN^{\phi
}(r_{+})\right) }{f^{\prime }(r_{+})}.  \label{phase_imaginary_Q}
\end{equation}

As the probabilities of crossing the horizon from inside to outside and
outside to inside is given by%
\begin{eqnarray}
P_{emission} &\varpropto &\exp \left( \frac{-2}{\hbar }\mathrm{Im}I\right)
=\exp \left( \frac{-2}{\hbar }(\mathrm{Im}W_{+}+\mathrm{Im}K)\right) , \\
P_{absorption} &\varpropto &\exp \left( \frac{-2}{\hbar }\mathrm{Im}I\right)
=\exp \left( \frac{-2}{\hbar }(\mathrm{Im}W_{-}+\mathrm{Im}K)\right) .
\end{eqnarray}%
The probability of a particle tunneling from inside to outside the horizon
is given by $\Gamma =\exp \left( \frac{-4}{\hbar }\mathrm{Im}W_{+}\right) $,
which on substituting the value of Im$W_{+}$ from equation (\ref%
{phase_imaginary_Q}) yields%
\begin{equation}
\Gamma =\exp \left( \frac{-4\pi }{\hbar }\frac{\left( \omega -qQ\ln \left(
\frac{r_{+}}{l}\right) -j\frac{J}{2r_{+}^{2}}\right) }{f^{\prime }(r_{+})}%
\right) .  \label{tunneling_c_r}
\end{equation}%
This is the tunneling probability of scalars across the event horizon of the
charged rotating BTZ black hole. We note that this does not depend on the
mass of the tunneling particle but only on its charge. By comparing this
with the Boltzmann factor of energy of particle, we can find the Hawking
temperature of this black hole $T_{H}=f^{\prime }(r_{+})/4\pi $ as
\begin{equation}
T_{H}=\frac{1}{4\pi }\left( \frac{2r_{+}}{l^{2}}-\frac{Q^{2}}{2r_{+}}-\frac{%
J^{2}}{2r_{+}^{3}}\right) .  \label{T_tunneling_Q_J}
\end{equation}%
Putting charge $Q=0$ will correspond to the temperature for
uncharged BTZ black hole \cite{Li:2008ws}.

Now we discuss the thermodynamic relations. The mass of the black hole is%
\begin{equation}
M_{bh}(Q,J)=\frac{r_{+}^{2}}{l^{2}}-\frac{1}{2}Q^{2}\ln \left( \frac{r_{+}}{l%
}\right) +\frac{J^{2}}{4r_{+}^{2}}.
\end{equation}%
The entropy is%
\begin{equation}
S=4\pi r^{+},
\end{equation}%
which is the circumference, in the unit $8G=1$. We see that%
\begin{equation}
\frac{\partial M}{\partial J}|_{Q,S}=\Omega =\frac{J}{2r_{+}^{2}},~~~~\ \
~~~J=2\Omega r_{+}^{2}.
\end{equation}%
The mass of the black hole can also be expressed as%
\begin{equation}
M=\left( \frac{1}{l^{2}}+\Omega ^{2}\right) r_{+}^{2}-\frac{1}{2}Q^{2}\ln
\left( \frac{r_{+}}{l}\right) .
\end{equation}%
For the electric potential $V$ we have
\begin{equation}
\frac{\partial M}{\partial Q}|_{J,S}=V=-Q\ln \left( \frac{r_{+}}{l}\right)
=A_{t}.
\end{equation}%
So that we have
\begin{equation}
\frac{\partial M}{\partial S}|_{Q,J}=T=\frac{1}{4\pi }\left( \frac{2r_{+}}{%
l^{2}}-\frac{Q^{2}}{2r_{+}}-\frac{J^{2}}{2r_{+}^{3}}\right) ,
\end{equation}%
which is from the thermodynamic relation, and is the same as obtained by the
quantum tunneling method above. The $J=0$ limit of the temperature reduces
to the result given by the quantum tunneling method in Section \ref{sec_
charge_scalar_charge}, and by another method \cite{Medved:2001ca}.

%%%%%%%%%%%%%%%%%%%%%%%%%%%%%%%%%%%%%%%%%%%%%%%%%%%%%%%%%%%%%%%%%%%
%%%%%%%%%%%%%%%%%%%%%%%%%%%%%%%%%%%%%%%%%%%%%%%%%%%%%%%%%%%%%%%%%%%
%%%%%%%%%%%%%%%%%%%%%%%%%%%%%%%%%%%%%%%%%%%%%%%%%%%%%%%%%%%%%%%%%%%

\section{Quantum tunneling of fermionic particles from charged BTZ black
holes}

\label{sec_ charge_fermionic_charge}

We will now calculate Dirac particle's Hawking radiation from the charged
BTZ black hole. In this case the function $f(r)~$will be, as in (\ref{bh_Q})
and (\ref{f_r_Q_}),
\begin{equation}
f(r)=-M+\frac{r^{2}}{l^{2}}-\frac{1}{2}Q^{2}\ln \left( \frac{r}{l}\right) .
\label{f_r_Q}
\end{equation}

We consider the two-component massive spinor field $\psi $, with mass $\mu $
and charge $q$, which obeys the covariant Dirac equation
\begin{equation}
i\hbar \gamma ^{a}e_{a}^{\mu }\left( \nabla _{\mu }-\frac{i}{\hbar }qA_{\mu
}\right) \psi -\mu \psi =0,
\end{equation}%
where $\nabla _{\mu }$ is the spinor covariant derivative given by $\nabla
_{\mu }=\partial _{\mu }+\Omega _{\mu }$,~where%
\begin{eqnarray}
\Omega _{\mu } &=&\frac{i}{2}\Gamma _{\mu }^{\alpha \beta }\Sigma _{\alpha
\beta }, \\
\Sigma _{\alpha \beta } &=&\frac{i}{4}\left[ \gamma ^{\alpha },\gamma
^{\beta }\right] ,~~~~~\Omega _{\mu }=\frac{-1}{8}\Gamma _{\mu }^{\alpha
\beta }\left[ \gamma ^{\alpha },\gamma ^{\beta }\right] .
\end{eqnarray}%
The $\gamma $ matrices in three spacetime dimensions are selected to be%
\begin{equation}
\gamma ^{a}=\left( -i\sigma ^{1},\sigma ^{0},\sigma ^{2}\right) ,
\end{equation}%
where $\sigma ^{i}$ are the Pauli sigma matrices. For the line element (\ref%
{bh_Q}) the vielbein field $e_{a}^{\mu }$ can be selected as
\begin{eqnarray}
e_{0}^{\mu } &=&\left(
\begin{array}{ccc}
f^{-\frac{1}{2}} & 0 & 0%
\end{array}%
\right) ,  \nonumber \\
e_{1}^{\mu } &=&\left(
\begin{array}{ccc}
0 & f^{\frac{1}{2}} & 0%
\end{array}%
\right) ,  \nonumber \\
e_{2}^{\mu } &=&\left(
\begin{array}{ccc}
0 & 0 & \frac{1}{r}%
\end{array}%
\right) .
\end{eqnarray}%
We use the ansatz for the two-component spinor $\psi $ as
\begin{equation}
\psi =\left(
\begin{array}{c}
A\left( t,r,\phi \right)  \\
B\left( t,r,\phi \right)
\end{array}%
\right) e^{\frac{i}{\hslash }I\left( t,\gamma ,\phi \right) }.
\end{equation}%
In order to apply WKB approximation, we insert ansatz for spinor
field $\psi $ into the Dirac equation. Dividing by the exponential
term with $\hbar$,
we have the following two equations%
\begin{equation}
A\left( \mu +\frac{1}{r}\partial _{\phi }I\left( t,r,\phi \right) \right) +B%
\left[ \sqrt{f}\partial _{r}I\left( t,r,\phi \right) +\left( \frac{1}{\sqrt{f%
}}\partial _{t}I\left( t,r,\phi \right) +\frac{1}{\sqrt{f}}Qq\ln \left(
\frac{r}{l}\right) \right) \right] =0,
\end{equation}%
\begin{equation}
A\left[ \sqrt{f}\partial _{r}I\left( t,r,\phi \right) -\left( \frac{1}{\sqrt{%
f}}\partial _{t}I\left( t,r,\phi \right) +\frac{1}{\sqrt{f}}Qq\ln \left(
\frac{r}{l}\right) \right) \right] +B\left( \mu -\frac{1}{r}\partial _{\phi
}I\left( t,r,\phi \right) \right) =0.
\end{equation}%
Note that although $A$ and $B$ are not constant, their derivatives and the
component $\Omega _{\mu }$ are all of order $\hbar ,$ so they can be
neglected to the lowest order in WKB approximation. These two equations have
a non-trivial solution for $A$ and $B$ if and only if the determinant of
coefficient matrix vanishes. Thus we get
\begin{equation}
\frac{1}{r^{2}}\left( \partial _{\phi }I\left( t,r,\phi \right)
\right) ^{2}+\mu ^{2}+\left( \sqrt{f}\partial _{r}I\left( t,r,\phi
\right) \right) ^{2}-\left( \frac{1}{\sqrt{f}}\partial _{t}I\left(
t,r,\phi \right) +\frac{1}{\sqrt{f}}Qq\ln \left(
\frac{r}{l}\right) \right) ^{2}=0.  \label{wave__}
\end{equation}

Because there are two Killing vectors $\left( \frac{\partial }{\partial t}%
\right) ^{\mu }$ and $\left( \frac{\partial }{\partial \phi }\right) ^{\mu }$
in the charged BTZ spacetime, so we can make the separation of variables for
$I\left( t,r,\phi \right) $ as
\begin{equation}
I=-\omega t+j\phi +W\left( r\right) +K,
\end{equation}%
where $\omega $ and $j$ are Dirac particle's energy and angular momentum
respectively, and $K$ is a complex constant. Now putting
\begin{equation}
\partial _{r}I=\partial _{r}W\left( r\right) ,~~~\partial _{\phi
}I=j,~~~\partial _{t}I=-\omega ,
\end{equation}%
in (\ref{wave__}) we get
\begin{equation}
\partial _{r}W\left( r\right) =\pm \frac{1}{f}\sqrt{\left( \omega -qQ\ln
\left( \frac{r}{l}\right) \right) ^{2}-f\left( \mu ^{2}+\frac{j^{2}}{r^{2}}%
\right) }.
\end{equation}%
In view of (\ref{p_emit_}) and (\ref{p_absorb_}), we have that $W_{-}=-W_{+}$%
. Integration gives
\begin{equation}
W_{+}\left( r\right) =\int \frac{dr}{f}\sqrt{\left( \omega -qQ\ln \left(
\frac{r}{l}\right) \right) ^{2}-f\left( \mu ^{2}+\frac{j^{2}}{r^{2}}\right) }%
.  \label{phase_r__}
\end{equation}%
Substituting the imaginary part of $W_{+}$ in tunneling probability we
obtain
\begin{equation}
W_{+}=\frac{\pi i}{f^{\prime }(r_{+})}\left( \omega -qQ\ln \left( \frac{r_{+}%
}{l}\right) \right) ,
\end{equation}%
or in simplified form we have
\begin{equation}
\mathrm{Im}W_{+}=\frac{\pi }{2\kappa }\left( \omega -qQ\ln \left( \frac{r_{+}%
}{l}\right) \right) ,
\end{equation}%
where $\kappa =\left( \frac{r_{+}}{l^{2}}-\frac{2GQ^{2}}{r_{+}}\right) $ is
the surface gravity of outer event horizon. This leads to the tunneling
probability%
\begin{equation}
\Gamma =\exp \left[ -\frac{2\pi }{\hbar \kappa }\left( \omega -qQ\ln (\frac{%
r_{+}}{l})\right) \right] .
\end{equation}%
Thus the Hawking temperature $T_{H}=f^{\prime }(r_{+})/4\pi $ is
\begin{equation}
T_{H}=\frac{r_{+}}{2\pi l^{2}}-\frac{Q^{2}}{8\pi r_{+}}.
\end{equation}%
This is the same as calculated in the case of scalar particles in (\ref%
{T_tunneling_Q}) in Section \ref{sec_ charge_scalar_charge}, and agrees with
the thermodynamic relation.

%%%%%%%%%%%%%%%%%%%%%%%%%%%%%%%%%%%%%%%%%%%%%%%%%%%%%%%%%%%%%%%%%%%
%%%%%%%%%%%%%%%%%%%%%%%%%%%%%%%%%%%%%%%%%%%%%%%%%%%%%%%%%%%%%%%%%%%
%%%%%%%%%%%%%%%%%%%%%%%%%%%%%%%%%%%%%%%%%%%%%%%%%%%%%%%%%%%%%%%%%%%

\section{Quantum tunneling of fermionic particles from rotating charged BTZ
black holes}

\label{sec_ charge_fermionic_charge_rotation}

In this section we work out the tunneling probability of fermions from
rotating charged BTZ black hole. We consider the Dirac equation for
electromagnetic field
\begin{equation}
i\gamma ^{\mu }\left( \partial _{\mu }+\Omega _{\mu }-\frac{i}{\hbar }%
qA_{\mu }\right) \psi -\frac{\mu }{\hbar }\psi =0,
\end{equation}
where
\begin{eqnarray}
\Omega _{\mu } &=&\frac{i}{2}\Gamma _{\mu }^{\alpha \beta }\Sigma _{\alpha
\beta }, \\
\Sigma _{\alpha \beta } &=&\frac{i}{4}\left[ \gamma ^{\alpha },\gamma
^{\beta }\right] ,~~~~\Omega _{\mu }=\frac{-1}{8}\Gamma _{\mu }^{\alpha
\beta }\left[ \gamma ^{\alpha },\gamma ^{\beta }\right] .
\end{eqnarray}
With the Pauli sigma matrices given by
\begin{equation}
\sigma ^{0}=\left(
\begin{array}{cc}
0 & 1 \\
1 & 0%
\end{array}%
\right) ,\sigma ^{1}=\left(
\begin{array}{cc}
0 & -i \\
i & 0%
\end{array}%
\right) ,\sigma ^{2}=\left(
\begin{array}{cc}
1 & 0 \\
0 & -1%
\end{array}%
\right) ,
\end{equation}%
we choose the curved space $\gamma ^{\mu }$ matrices as
\begin{eqnarray}
\gamma ^{t} &=&\left(
\begin{array}{cc}
0 & -\frac{1}{\sqrt{f}} \\
\frac{1}{\sqrt{f}} & 0%
\end{array}%
\right) ,\gamma ^{r}=\left(
\begin{array}{cc}
0 & \sqrt{f} \\
\sqrt{f} & 0%
\end{array}%
\right) ,  \nonumber \\
\gamma ^{\phi } &=&\left(
\begin{array}{cc}
\frac{1}{r} & -\frac{4JG}{r^{2}\sqrt{f}} \\
\frac{4JG}{r^{2}\sqrt{f}} & -\frac{1}{r}%
\end{array}%
\right) ,
\end{eqnarray}%
which also satisfy the condition $\left\{ \gamma ^{\mu },\gamma ^{\nu
}\right\} =2g^{\mu \nu }\times ~$\textrm{I} where I is the identity matrix.

Now inserting the value of electric potential for charged BTZ black hole the
equation of motion becomes
\begin{equation}
i\gamma ^{t}\left( \partial _{t}-\frac{i}{\hbar }qA_{t}\right) \psi +i\gamma
^{r}\left( \partial _{r}\right) \psi +i\gamma ^{\phi }\left( \partial _{\phi
}\right) \psi -\frac{\mu }{\hbar }\psi =0.  \label{Dirac_}
\end{equation}%
For a fermion with spin 1/2 the wave function has two states namely spin-up (%
$\uparrow $) and spin-down ($\downarrow $), and therefore, we can
take the following ansatz for the solution
\begin{eqnarray}
\psi _{\uparrow } &=&\left(
\begin{array}{c}
A\left( t,r,\phi \right)  \\
0%
\end{array}%
\right) e^{\frac{i}{\hslash }I_{\uparrow }\left( t,r,\phi \right) },
\label{spin_up_} \\
\psi _{\downarrow } &=&\left(
\begin{array}{c}
0 \\
B\left( t,r,\phi \right)
\end{array}%
\right) e^{\frac{i}{\hslash }I_{\downarrow }\left( t,r,\phi \right) },
\label{spin_down_}
\end{eqnarray}%
where $\psi _{\uparrow }$ denotes the wave function of the spin-up
particle and $\psi _{\downarrow }$ is for the spin-down case. Inserting equation (\ref%
{spin_up_}) for the spin-up particle into the Dirac equation
(\ref{Dirac_}) and dividing by exponential term and multiplying by
$\hbar $, we get the following equation
\begin{equation}
-\frac{A}{\sqrt{f}}\partial _{t}I_{\uparrow }\left( t,r,\phi \right) +\frac{%
qA_{t}}{\sqrt{f}}A-\sqrt{f}A\partial _{r}I_{\uparrow }\left( t,r,\phi
\right) -\frac{4JGA}{r^{2}\sqrt{f}}\partial _{\phi }I_{\uparrow }\left(
t,r,\phi \right) =0.
\end{equation}%
Now considering the method of separation of variables for the
spin-up case we have
\begin{equation}
I_{\uparrow }=-\omega t+W(r)+\Theta (\phi )+K=-\omega t+W(r)+j\phi +K.
\label{spin_up_phase}
\end{equation}%
Here $\omega $ and $j$ are the energy and angular momentum of the emitted
particle, and $K$ is a complex constant. Using this expression in the above
equation we get
\begin{equation}
\frac{A}{\sqrt{f}}\omega +\frac{qA_{t}}{\sqrt{f}}A-\sqrt{f}A\partial _{r}W-%
\frac{4JGA}{r^{2}\sqrt{f}}\partial _{\phi }\Theta =0.  \label{eqn_spin_}
\end{equation}%
For simplification we put equation (\ref{spin_up_phase}) in (\ref{eqn_spin_}%
) to get
\begin{equation}
\frac{\omega }{\sqrt{f}}+\frac{qA_{t}}{\sqrt{f}}-\sqrt{f}\partial _{r}W-j%
\frac{4JG}{r^{2}\sqrt{f}}=0,
\end{equation}%
or
\begin{equation}
\partial _{r}W=\frac{1}{f}\left( \omega +qA_{t}-j\frac{4JG}{r^{2}}\right) .
\label{spin_up_w}
\end{equation}

If we look at the spin-down particle, its phase $I_{\downarrow }$
and its $r$-dependence have the similar expressions as equations (\ref%
{spin_up_phase}) and (\ref{spin_up_w}), respectively. Integration of
equation (\ref{spin_up_w}) gives
\begin{equation}
W=\int \frac{dr}{f}\left( \omega +qA_{t}-j\frac{4JG}{r^{2}}\right) .
\end{equation}%
We integrate along a semi circle around the pole at $r_{+}=0.$ Now, at the
horizon the radial function can be given as%
\begin{equation}
W=\frac{\pi i(\omega +qA_{t}-j\frac{4JG}{r_{+}^{2}})}{\left( \frac{2r_{+}}{%
l^{2}}-\frac{4GQ^{2}}{r_{+}}-\frac{32G^{2}J^{2}}{r_{+}^{3}}\right) }.
\end{equation}%
Tunneling probability for this is given by equation (\ref{rate_phase_}) that
is
\begin{equation}
\Gamma =\exp \left( \frac{-\pi (\omega -qQ\ln \left( \frac{r_{+}}{l}\right)
-j\frac{4JG}{r_{+}^{2}})}{\left( \frac{r_{+}}{2l^{2}}-\frac{GQ^{2}}{r_{+}}-%
\frac{8G^{2}J^{2}}{r_{+}^{3}}\right) \hbar }\right) ,
\label{tunneling_c_r_f}
\end{equation}%
where $\kappa =\left( \frac{r_{+}}{l^{2}}-\frac{2GQ^{2}}{r_{+}}-\frac{%
16G^{2}J^{2}}{r_{+}^{3}}\right) $ is the surface gravity. We work in the
units $8G=1$. We see that this is the same as obtained by solving the
Klein-Gordon equation in Section \ref{sec_ charge_scalar_charge_rotation}.

%%%%%%%%%%%%%%%%%%%%%%%%%%%%%%%%%%%%%%%%%%%%%%%%%%%%%%%%%%%%%%%%%%%
%%%%%%%%%%%%%%%%%%%%%%%%%%%%%%%%%%%%%%%%%%%%%%%%%%%%%%%%%%%%%%%%%%%
%%%%%%%%%%%%%%%%%%%%%%%%%%%%%%%%%%%%%%%%%%%%%%%%%%%%%%%%%%%%%%%%%%%

\section{Quantum tunneling from three-dimensional topological black holes}

\label{sec_ particle_sol}

Some classes of three-manifolds have been studied for their interesting
physical properties. One such example is the Sol geometry \cite{Scott}
defined by
\begin{equation}
ds^{2}=e^{2u}dx^{2}+du^{2}+e^{-2u}dy^{2}.  \label{Sol_def_}
\end{equation}%
It is the \textrm{R}$^{2}~$bundle over \textrm{R},~in which ($x,y$) is the
\textrm{R}$^{2}$, and $u$ is the \textrm{R}. From this class we consider the
following metric representing a three-dimensional topological black hole
\begin{equation}
ds^{2}=\frac{-f}{Ml^{2}}dt^{2}+\frac{1}{l^{2}f}dr^{2}+\frac{M}{f}d\phi ^{2},
\label{bh_Sol_01}
\end{equation}%
where
\begin{equation}
f(r)=\frac{r^{2}}{l^{2}}-M.  \label{sol_f}
\end{equation}%
In the asymptotic region, when $f(r)\rightarrow r^{2}/l^{2}$, the above
metric becomes%
\begin{equation}
ds^{2}=\frac{-r^{2}}{Ml^{4}}dt^{2}+\frac{dr^{2}}{r^{2}}+\frac{Ml^{2}}{r^{2}}%
d\phi ^{2}.
\end{equation}%
We redefine
\begin{equation}
\tilde{t}=\frac{t}{\sqrt{M}}=l^{2}x,~~~~~~r=e^{u},~~~~u=\ln r,~~~~\ \ y=%
\sqrt{M}l\phi .~~
\end{equation}%
In this case (\ref{bh_Sol_01}) is asymptotic to
\begin{eqnarray}
ds^{2} &=&-\frac{f}{l^{2}}d\tilde{t}^{2}+\frac{1}{l^{2}f}dr^{2}+\frac{M}{f}%
d\phi ^{2} \\
&\rightarrow &-r^{2}dx^{2}+\frac{dr^{2}}{r^{2}}+\frac{1}{r^{2}}dy^{2} \\
&=&-e^{2u}dx^{2}+du^{2}+e^{-2u}dy^{2}.  \label{Sol_lorentzian}
\end{eqnarray}%
We see that (\ref{Sol_lorentzian}) is exactly the Lorentzian
version of the Sol geometry (\ref{Sol_def_}), by analytical
continuation $x\rightarrow ix$. So the black hole in
(\ref{bh_Sol_01}) is asymptotic to the Sol geometry. The metric
(\ref{bh_Sol_01}) on three-manifolds can be interpreted as black
holes or black hole like objects. They can arise from three
dimensional gravity with matter fields.

Here we study the emission of particles from these Sol black holes described
above. We use the Dirac equation
\begin{equation}
i\hbar \gamma ^{a}e_{a}^{\mu }\tilde{\nabla}_{\mu }\psi -\mu \psi =0,
\end{equation}%
where $\tilde{\nabla}_{\mu }$ is the spinor covariant derivative. It is
worth mentioning here that using Klein-Gordon equation will also give the
same results. We select the $\gamma $ matrices as before and write the
vielbein field $e_{a}^{\mu }$ as
\begin{eqnarray}
&&e_{0}^{\mu }=\left(
\begin{array}{ccc}
\frac{M^{1/2}l}{f^{1/2}} & 0 & 0%
\end{array}%
\right) ,  \nonumber \\
&&e_{1}^{\mu }=\left(
\begin{array}{ccc}
0 & lf^{1/2} & 0%
\end{array}%
\right) ,  \nonumber \\
&&e_{2}^{\mu }=\left(
\begin{array}{ccc}
0 & 0 & \frac{f^{1/2}}{M^{1/2}}%
\end{array}%
\right) .
\end{eqnarray}%
We use the ansatz for the two-component spinor $\psi $ as
\begin{equation}
\psi =\left(
\begin{array}{c}
A\left( t,r,\phi \right)  \\
B\left( t,r,\phi \right)
\end{array}%
\right) e^{\frac{i}{\hslash }I\left( t,\gamma ,\phi \right) }.
\end{equation}%
In order to apply WKB approximation, we insert this ansatz for spinor field $%
\psi $ into the Dirac equation. Dividing by the exponential term
with $\hbar $, one can get the following two equations
\begin{eqnarray}
A\left( \mu +\sqrt{\frac{f}{M}}\partial _{\phi }I\left( t,r,\phi \right)
\right) +B\left( \sqrt{f}l\partial _{r}I\left( t,r,\phi \right) +\sqrt{\frac{%
M}{f}}l\partial _{t}I\left( t,r,\phi \right) \right)  &=&0,  \nonumber \\
&& \\
A\left( \sqrt{f}l\partial _{r}I\left( t,r,\phi \right) -\sqrt{\frac{M}{f}}%
l\partial _{t}I\left( t,r,\phi \right) \right) +B\left( \mu -\sqrt{\frac{f}{M%
}}\partial _{\phi }I\left( t,r,\phi \right) \right)  &=&0.  \nonumber \\
&&
\end{eqnarray}%
Now as we have discussed in the previous sections, for nontrivial solution
we put the determinant of coefficient matrix equal to zero
\begin{equation}
\frac{M}{f}l^{2}\left( \partial _{t}I\right) ^{2}-\frac{f}{M}\left( \partial
_{\phi }I\right) ^{2}-l^{2}f\left( \partial _{r}I\right) ^{2}-\mu ^{2}=0.
\label{wave_}
\end{equation}%
To calculate the classical action of trajectory we use the method of
separation of variables and suppose%
\begin{eqnarray}
I &=&j\phi -\omega \tilde{t}+W(r)+K \\
&=&j\phi -\frac{1}{\sqrt{M}}\omega t+W(r)+K,
\end{eqnarray}%
where $K$ is a complex constant and $j$ is the angular momentum of the
particle. Now we put the relations
\begin{equation}
\partial _{\phi }I=j,~~\partial _{t}I=-\frac{\omega }{\sqrt{M}},~~\partial
_{r}I=\partial _{r}W,
\end{equation}%
in equation (\ref{wave_}) and obtain%
\begin{equation}
\partial _{r}W(r)=\frac{1}{f}\sqrt{\omega ^{2}-f\left( \frac{\mu }{l}\right)
^{2}-\frac{f^{2}}{M}\left( \frac{j}{l}\right) ^{2}}.
\end{equation}%
So we see that%
\begin{eqnarray}
W &=&\int dr\frac{1}{f}\sqrt{\omega ^{2}-f\left( \frac{\mu }{l}\right) ^{2}-%
\frac{f^{2}}{M}\left( \frac{j}{l}\right) ^{2}} \\
&=&\frac{\pi i(\omega )}{f^{\prime }(r_{+})},  \label{R_}
\end{eqnarray}%
where we have integrated along the semi-circle around the pole for $f=0$~at $%
r_{+}$.

This implies that the probability of a particle tunneling from
inside to outside the horizon becomes
\begin{equation}
\Gamma =\exp \left( -\frac{4}{\hbar }\mathrm{Im}W\right) =\exp \left( -\frac{%
4\pi \omega }{f^{\prime }(r_{+})\hbar }\right) .
\end{equation}%
Comparing with the Boltzmann factor $\exp \left( -\beta \omega \right) $ we
obtain the temperature $T=f^{\prime }(r_{+})/4\pi $. Substituting the
expression of $f(r)$, for $r_{+}=\sqrt{M}l$,
\begin{equation}
T=\frac{f^{\prime }(r_{+})}{4\pi }=~\frac{r_{+}}{2\pi l^{2}}=\frac{\sqrt{M}}{%
2\pi l}.
\end{equation}%
From the thermodynamic relation%
\begin{equation}
\frac{\partial M}{\partial S}=T,  \label{M_T}
\end{equation}%
the entropy $S$ is%
\begin{equation}
S=4\pi r^{+}=4\pi \sqrt{M}l.  \label{Sol_S}
\end{equation}%
The temperature and thermodynamic behavior are similar to those of BTZ black
hole.

\section{Discussion}

The theory of three-manifolds has been studied for its interesting geometric
and topological properties \cite{Scott}, and classifications on one hand,
and its physical applications on the other. Some of the metrics on
three-manifolds have been interpreted as black holes or black hole like
objects. They play a crucial role in understanding lower dimensional gravity
theories.

Here we have analyzed the quantum tunneling approach of Hawking
radiations for three-dimensional BTZ and other topological black
holes. The charged BTZ black hole represents a solution of the
Einstein-Maxwell equations. When we include the charge it becomes
a three-dimensional analogue of the Reissner-Nordstr\"{o}m black
hole. We also computed the quantum tunneling for the topological
black holes asymptotic to Sol geometry. Laws of black hole
mechanics can be extended to these objects as well.

The BTZ spacetime also appears from the near horizon geometry of
higher dimensional black holes. In the near horizon geometry of
higher dimensional black holes, under appropriate limit, the time
and radial direction of the black hole, and a circle direction in
the extra dimensions, combine into a BTZ geometry
\cite{Balasubramanian:2007bs, Fareghbal:2008ar}. It is interesting
to see the connection of this to the discussion in this paper.

We have studied the emission of charged scalar particles and fermions from
charged BTZ black holes, with and without rotation. For this purpose we have
solved the charged Klein-Gordon and Dirac equations using WKB approximation
and symmetries of the background spacetime. Using complex path integration
we worked out the tunneling probability of particles from charged BTZ black
holes. This also yields Hawking temperature from these three dimensional
objects. The temperature is given by $f^{\prime }(r_{+})/4\pi $, where the
function $f(r)$ for the charged black hole is given by equation (\ref{f_r_Q_}%
) and that for the charged rotating black hole is given by
(\ref{f_r_Q__}). The charged BTZ black holes have a -$\ln
r~$potential term which is very interesting. We also obtained the
Boltzmann factors with chemical potentials conjugate to the charge
and to the angular momentum of the particles. It is worth
emphasizing here that for no value of the energy, charge and
angular momentum of the particles will the tunneling probabilities
be greater than 1. Thus they will not violate unitarity. This is
taken care of by the temporal contribution to the imaginary part
of the action \cite{Akhmedov:2008ru, Akhmedova:2008dz}.

A three-dimensional topological black hole which is asymptotic to a Sol
three-manifold has also been investigated for its thermodynamical
properties. We have studied the emission of particles and find that the
Hawking temperature is again given by $f^{\prime }(r_{+})/4\pi $, for $f(r)$
given by (\ref{sol_f}), and it goes as the square root of $M$. These objects
have some other interesting mathematical and physical properties as well.

\acknowledgments

A grant from the Council for International Exchange of Scholars (CIES),
Washington, DC, under the Fulbright Fellowship program is gratefully
acknowledged. This work was supported in part by NSF grant DMS-1159412, NSF
grant PHY-0937443, NSF grant DMS-0804454, and by the Fundamental Laws
Initiative of the Center for the Fundamental Laws of Nature, Harvard
University. We thank R. de Mello Koch, S. Ramgoolam, J. Heckman, Y. Kimura
for correspondence.

\end{document}